\documentclass[%
 reprint,
 amsmath,amssymb,
prb,
]{revtex4-2}

\pdfoutput=1
\usepackage{multirow}
\usepackage{graphicx}
\usepackage{dcolumn}
\usepackage{bm}
\usepackage{hyperref}
\usepackage[mathlines]{lineno}
\usepackage{float}

\begin{document}

\preprint{APS/123-QED}

\title{Bilayer stacking A-type altermagnet: A general approach to generating two-dimensional altermagnetism}

\author{Sike Zeng}

\author{Yu-Jun Zhao}%
\email{zhaoyj@scut.edu.cn}
\affiliation{%
	Department of Physics, South China University of Technology, Guangzhou 510640, China
}%

\date{\today}

\begin{abstract}
 In this article, we propose a new concept of bilayer stacking A-type altermagnet (BSAA), in which two identical ferromagnetic monolayers are stacked with antiferromagnetic coupling to form a two-dimensional A-type altermagnet. By solving the stacking model, we derive all  BSAAs for all layer groups and draw three key conclusions: (\romannumeral1) Only 17 layer groups can realize intrinsic A-type altermagnetism. All 2D A-type altermagnets must belong to these 17 layer groups, which will be helpful to search for 2D A-type altermagnet. (\romannumeral2) It is impossible to connect the two sublattices of BSAA using $S_{3z}$ or $S_{6z}$, a constraint that is also applicable to all 2D altermagnets. (\romannumeral3) $C_{2\alpha}$ is a general stacking operation to generate BSAA for an arbitrary monolayer. Our theory not only can explain the previously reported twisted-bilayer altermagnets, but also can provide more possibilities to generate A-type altermagnets. Our research has significantly broadened the range of candidate materials for 2D altermagnets. Based on conclusion (\romannumeral1), the bilayer NiZrCl$_6$ is predicted to exhibit intrinsic A-type altermagnetism. Additionally, we use twisted-bilayer NiCl$_2$, previously reported in the literature, as the second example of BSAA. Furthermore, utilizing symmetry analysis and first-principles calculation, we scrutinize their spin-momentum locking characteristic to substantiate their altermagnetic properties.

\end{abstract}

\maketitle

\section{\label{sec:level1} INTRODUCTION}
Altermagnetism, characterized by collinear-compensated magnetic order in real space and time-reversal symmetry breaking in reciprocal space, has recently attracted substantial interest within the field of condensed matter physics\cite{mazin2024altermagnetism,PhysRevX.12.040501,bai2024altermagnetism}, due to its intriguing physical properties and promising application in spintronics. Initially proposed in several scientific literature\cite{vsmejkal2020crystal,hayami2019momentum,PhysRevB.102.014422,mazin2021prediction,ma2021multifunctional} and subsequently formalized under the term `altermagnetism'\cite{vsmejkal2022beyond}, this phenomenon has emerged as a novel third magnetic state within the context of spin group. Unlike  antiferromagnet, altermagnet is characterized by the connection of opposite sublattices via proper or improper rotation, rather than translation or inversion symmetry, leading to the disruption of \textbf{\emph{PT}} symmetry\cite{vsmejkal2022beyond}. Recently, the spin-splitting electronic structure has been reported experimentally in MnTe, utilizing angle-resolved photoemission spectroscopy measures, which provides direct evidence for altermagnetism\cite{lee2024broken}. From an application perspective, altermagnet presents considerable potential for applications in spintronics,  as exhibiting the advantage of no stray field, THz spin dynamics and strong time-reversal symmetry breaking. It has been suggested that giant/tunneling magnetoresistance (GMR/TMR) effect\cite{PhysRevX.12.011028} can be generated in altermagnets. Moreover, as a new type of torque, spin-splitter torque (SST), combining the advantage of spin-transfer torque (STT) and spin-orbit torque (SOT), has been reported in the altermagnet theoretically\cite{PhysRevLett.126.127701} and experimentally\cite{PhysRevLett.129.137201,PhysRevLett.128.197202}.

To date, a multitude of three-dimensional (3D) altermagnets have been identified\cite{bai2024altermagnetism}. However, two-dimensional (2D) intrinsic altermagnets were rarely reported\cite{sodequist2024two,zeng2024descriptiontwodimensionalaltermagnetism}. The expansion of two-dimensional materials exhibiting altermagnetism will greatly enhance our understanding of the fundamental properties of altermagnets and promote their application in the spintronics. Therefore, significant efforts have been devoted to enriching the candidate materials for 2D altermagnet\cite{mazin2023induced,https://doi.org/10.1002/adfm.202402080,PhysRevMaterials.8.L051401,PhysRevLett.130.046401,liu2024twistedmagneticvander}. Among these, it is particularly noteworthy that the realization of altermagnetism through twisting in a bilayer without altermagnetism offers a general approach to generating two-dimensional altermagnet, substantially broadening the spectrum of candidate materials for 2D altermagnet\cite{PhysRevMaterials.8.L051401,PhysRevLett.130.046401,liu2024twistedmagneticvander}. 

In 2D systems, the spin degeneracy of the nonrelativistic band structure for all \textbf{k}-vectors in the whole Brillouin zone is protected by four spin-group symmetries\cite{zeng2024descriptiontwodimensionalaltermagnetism}. The spin-group symmetry is denoted as $[R_i||R_j]$, where the transformations on the left of the double bar only act on the spin space and those on the right of the double bar only act on the real space. For all 2D systems mentioned in this article, it is assumed that they lie parallel to the \emph{xy}-plane. Firstly, $[C_2||\tau]$ symmetry, where $C_2$ represents 2-fold rotation around an axis perpendicular to the spins and $\tau$ represents translation symmetry connecting atoms of the opposite sublattices, results in the spin-degeneracy band structure. Moreover, the spin degeneracy is also protected by $[C_2||\overline{E}]$, in which $\overline{E}$ represents inversion symmetry. Its physical signification is that the atoms of the opposite sublattices are connected by inversion symmetry. The aforementioned symmetries also maintain band structure spin-degeneracy in 3D systems. However, compared to 3D systems, there are two extra symmetries to protect spin degeneracy in 2D systems. One is $[C_2||m_z]$, which represents that the atoms of the opposite sublattices can be connected by the mirror symmetry through the  \emph{xy}-plane. The other is $[C_2||C_{2z}]$, which represents the atoms of the opposite sublattices can be connected by a 2-fold rotation around \emph{z}-axis. 

Based on the above discussion, we can conclude the requirements for the emergence of altermagnetism in 2D materials: (\romannumeral1) The system must exhibit collinear-compensated magnetic order, i.e., it must have two sublattices with antiparallel magnetic moments. (\romannumeral2) The atoms of the opposite sublattices cannot be connected by translation symmetry, inversion symmetry, mirror symmetry through the \emph{xy}-plane or a 2-fold rotation around \emph{z}-axis, given that the system is parallel to \emph{xy}-plane. (\romannumeral3) There must exist at least one symmetry operation that interconnects the opposite sublattices, which can be proper or improper rotation. Therefore, we can discern that the emergence of 2D altermagnets entails more requirement on symmetry, which may be why 2D intrinsic altermagnets were rarely reported.

In 2D systems, translational symmetry is inherently absent in the out-of-plane direction, which serves as the foundation for this work. A-type altermagnetism involves a bilayer system that can be considered as comprising two single ferromagnetic layers. In this system, the bottom layer remains stationary, while the top layer is stacked on top with antiferromagnetic coupling. We notice that A-type altermagnets, which exhibit intralayer ferromagnetic coupling and interlayer antiferromagnetic coupling, inherently satisfy the requirement that the opposite sublattices cannot be connected by translation symmetry and a 2-fold rotation around the \emph{z}-axis. Therefore, we believe that A-type altermagnetism is easier to appear in 2D materials. 

Here, we propose a new concept of bilayer stacking A-type altermagnet (BSAA). Utilizing the stacking model, we have solved for all BSAAs, as listed in Table \ref{table1}. Interestingly, based on the solutions, we draw three intriguing conclusions in Section \ref{sec:level2}. Moreover, we predict that bilayer NiZrCl$_6$ exhibits intrinsic A-type altermagnetism in Section \ref{sec:level3}, which has never been reported previously. Finally, we notice that twisted-bilayer A-type altermagnets reported in the previous work\cite{PhysRevLett.130.046401,liu2024twistedmagneticvander} is encompassed within BSAA and can be understood using this theory, as twisting can be regarded as a stacking operation. Therefore, we use twisted-bilayer NiCl$_2$, which has been reported in the previous work\cite{PhysRevLett.130.046401}, as the second example of BSAA in Section \ref{sec:level4}. Of course, these two examples are just the beginning of BSAA. We believe that there is a vast range of BSAA, as Table \ref{table1} presents a diverse set of stacking operations that can generate BSAA.
 
\section{\label{sec:level2} ALL POSSIBILITIES OF BILAYER STACKING A-TYPE ALTERMAGNETS}
We now focus on the derivation of all BSAAs for all layer groups, where two identical ferromagnetic monolayers are stacked with antiferromagnetic coupling to form a two-dimensional A-type altermagnet. Employing different stacking configurations, a specific  monolayer can form bilayers with different symmetry. As a result, some stacking configurations can generate A-type altermagnetism for a specific ferromagnetic monolayer, while others can not.

Firstly, we introduce the model for BSAA. The model is based on the sole assumption of 
intralayer ferromagnetic coupling and interlayer antiferromagnetic coupling, which satisfies the requirement (\romannumeral1) of forming 2D altermagnetism. Due to the nonrelativistic nature of altermagnetism, real space and spin space are independent of each other. Therefore, we focus only on real-space transformations, i.e., space-group symmetry, because of the above assumption. Meanwhile, we find that two remaining requirements, requirement (\romannumeral2) and (\romannumeral3), are solely related to the symmetry operations that interconnect two sublattices. Consequently, we are primarily concerned with the symmetry operations that exchange the positions of the two ferromagnetic layers. 

All 2D point-group symmetry operations can be summarized into seven types\cite{PhysRevLett.130.146801}:(\romannumeral1) identity $E$; (\romannumeral2) inversion $\overline{E}$; (\romannumeral3) mirror symmetry through \emph{xy}-plane $m_z$; (\romannumeral4) mirror symmetry, $m_\alpha$, is perpendicular to \emph{xy}-plane and parallel to the direction (cos($\alpha$), sin($\alpha$), 0); (\romannumeral5) \emph{n}-fold rotation symmetry along \emph{z}-axis $C_{nz}$; (\romannumeral6) 2-fold rotation symmetry, $C_{2\alpha}$, along the direction (cos($\alpha$), sin($\alpha$), 0); (\romannumeral7) \emph{n}-fold rotation symmetry with mirror symmetry along \emph{z}-axis $S_{nz}$. According to whether they reverse the \emph{z}-coordination, they can further be classified into two categories $R^+$ and $R^-$. $R^-$ can reverse the \emph{z}-coordination, while $R^+$ can not. $R^-$ contains $\{\overline{E},m_z,C_{2\alpha},S_{nz}\}$, and $R^+$ contains $\{E,C_{nz},m_\alpha\}$. In the context of this article, $R^-$ represents the symmetry operations that exchange the positions of the two ferromagnetic layers, while $R^+$ does not. Therefore, we conclude that to obtain all BSAAs for all layer groups, it is sufficient to calculate all stacking operations generating  $R^-$ in the bilayer for each layer group. Subsequently, we exclude the stacking operations generating $\{\overline{E}, m_z\}$ in the bilayer from those generating $\{C_{2\alpha}, S_{nz}\}$ in the bilayer for each layer group. This ensures that two single layers cannot be connected by $\overline{E}$ or $m_z$, but can be connected by $C_{2\alpha}$ or $S_{nz}$. Finally, the remaining stacking operations are all the possible operations to generate BSAA from a ferromagnetic monolayer.

Let us assume that the single ferromagnetic layer $S$, with layer group $G_s=\{\hat{R}_s\}$, is parallel to \emph{xy}-plane. In this article, symbols with a caret represent space-group symmetry, while those without a caret represent point-group symmetry. For example, $\hat{R}_s$ represents the space-group symmetry of layer $S$, while ${R}_s$ represents the point-group symmetry. The second ferromagnetic layer $S'$ is generated by applying the stacking operation $\hat{\tau}_z\hat{O}$ on S, where $\hat{\tau}_z$ is an arbitrary trivial translation operation along \emph{z} direction and cannot affect the symmetry of the bilayer B whose layer group is $G_B=\{\hat{R}_B\}$. $\hat{O}$ is the non-trivial part of stacking operation, and $\hat{O}=\{\tau_o|O\}$, where $\tau_o$ represents the translation part and $O$ is the point-group part. As a result, we can obtain the stacking model:
\begin{equation}
	\left\{\label{eq:1}
	\begin{aligned}
		S &= \hat{R}_sS, \ \ \forall \hat{R}_s \in G_s \\
		B &= \hat{R}_BB, \ \ \forall \hat{R}_B \in G_B\\
		B &= S+S' \\
		S'&= \hat{\tau}_z\hat{O} S
	\end{aligned}
	\right.
\end{equation}
Due to considering only real-space transformations, this model is analogous to that of the bilayer stacking ferroelectricity (BSF)\cite{PhysRevLett.130.146801}. Therefore, we directly use the derivation results of BSF. 

We only list the results about $R^-$, and more details are available in reference\cite{PhysRevLett.130.146801}. To ensure that the bilayer formed by stacking possesses the symmetry $R^{-}$ $(R^{-} = \overline{E}$, $m_z$  or $C_{2\alpha})$, the stacking operations must satisfy the following equations:
\begin{align}
	 O &\in R^{-} G_{s0}  \label{eq:2} \\
	(E + &R^{-}) \tau_{o} = \tau_{s0} \label{eq:3}
\end{align}
Equation (\ref{eq:2}) imposes a constraint on the point-group part of stacking operation, while Equation (\ref{eq:3}) impose a constraint on the translation part of stacking operation. $G_{s0}$ contains all point-group symmetry of $G_{s}$. $\tau_{s0}$ is an integer translation of the lattice vectors of the conventional cell of layer $S$. Moreover, to ensure that the bilayer possesses the symmetry $S_{nz}$, the stacking operations must satisfy the following equations:

 if $S_{nz} \in G_{s0}$: 
  \begin{align}
 	O &\in S_{nz} G_{s0}  \label{eq:4} \\
 	(E + &S_{nz}) \tau_{o} = \tau_{s0} \label{eq:5}
 \end{align}
 
 if $S_{nz} \notin G_{s0}$: 
\begin{align}
	&C_{nz}^2 \in G_{s0} \label{eq:6}\\
	&O \in S_{nz} G_{s0}  \label{eq:7} \\
	(E + &S_{nz}) \tau_{o} = \tau_s^+ + \tau_{s0} \label{eq:8}
\end{align}
Equation (\ref{eq:6}) imposes a constraint on $G_{s0}$. $\tau_s^+$ is the translation part of $\hat{C}_{nz}^2$ in group $G_{s}$. The detailed derivation from Equation (\ref{eq:1}) to Equations (\ref{eq:2})-(\ref{eq:8}) is available in reference\cite{PhysRevLett.130.146801}. However, the general solution of these equations for each layer group is not available in reference\cite{PhysRevLett.130.146801}. 

We now derive all stacking operations to generate BSAA for each layer group, based on these equations. Firstly, we notice that there must be $C_{nz}^2 \in G_{s0}$ when $S_{nz} \in G_{s0}$, and $\tau_s^+$ is zero for any $C_{nz}^2$. Therefore, Equations  (\ref{eq:2})-(\ref{eq:8}) can be unified into the following equations:

For  $R^{-}$ $(R^{-} = \overline{E}, m_z, S_{nz},C_{2\alpha})$:
\begin{align}
	O &\in R^{-} G_{s0}  \label{eq:9} \\
	(E + &R^{-}) \tau_{o} = \tau_{s0} \label{eq:10}
\end{align}

Additionally, for $S_{nz}$:
\begin{align}
	&C_{nz}^2 \in G_{s0} \label{eq:11}
\end{align}
Equation (\ref{eq:11}) determines whether it is possible for a specific monolayer to form the bilayer possessing $S_{nz}$ symmetry through stacking. In summary, to ensure that the bilayer possesses the symmetry $S_{nz}$, the point-group part of space-group symmetry of layer S must contain $C_{nz}^2$ and the stacking operations must satisfy Equations (\ref{eq:9}, \ref{eq:10}). For other symmetries, the requirement is that the stacking operations must satisfy Equations (\ref{eq:9}, \ref{eq:10}). Through observing Equations (\ref{eq:9})-(\ref{eq:11}), we find that these equations are independent of the translation part of $G_s$. Therefore, the 80 layer groups can be processed into 36 point groups $G_{s0}$. By applying the 36 point groups and the four possible $R^{-}$ symmetries into Equation (\ref{eq:9}), the point-group part of all stacking operations can be obtained for all layer groups.

 \squeezetable 
\renewcommand{\arraystretch}{1.4}
\begin{table*}
	\caption{\label{table1}All stacking operations that can construct BSAA from  ferromagnetic monolayer. The layer groups of monolayer are listed in the first column. The point group part of stacking operations for each group are listed in the second column. The third column lists the symmetry exchanging two ferromagnetic layers in the bilayer under the stacking operation listed in the second column. The translation part of stacking operations, $\tau_o$, is only dependent on the interlayer symmetry. For $C_{2\alpha}$, translation of arbitrary length perpendicular to the vector (cos($\alpha$), sin($\alpha$), 0), as well as translations by half of the lattice vector in the direction of vector (cos($\alpha$), sin($\alpha$), 0), and their combinations, are all permitted. For $S_{4z}$, the translation operation should satisfy Equation (\ref{eq:12}). In some cases, certain translations are forbidden, which have been listed in the second column. The symbols $E,C,m,S$ represent identity, rotation, mirror and rotation with mirror operation respectively. The plus (+) and minus (-)  denote the counterclockwise and clockwise rotations, respectively. The angle in the upper right corner of $S_z$ and $C_z$ represents the rotation angle. The angle in the upper right corner of $C_2$ and $\alpha$ represents the angle relative to \emph{x}-axis. The range of $\alpha$ is from 0 to $\pi$. $G=(0,0)$, $A=(\frac{1}{2},0)$, $B=(0,\frac{1}{2})$, $C=(\frac{1}{2},\frac{1}{2})$, $G'=(\frac{1}{4},\frac{1}{4})$, $A'=(\frac{3}{4},\frac{1}{4})$, $B'=(\frac{1}{4},\frac{3}{4})$, $G'=(\frac{3}{4},\frac{3}{4})$, $M=(\frac{1}{3},\frac{2}{3})$, $N=(\frac{2}{3},\frac{1}{3})$, employing lattice vectors as basis vectors.}
	
	\begin{ruledtabular}
		\begin{tabular}{ccc}
			\textrm{Layer group}&
			\textrm{point group part of stacking operation} &
			\textrm{interlayer symmetry}\\
			\hline
			1,49,65&$C_{2\alpha}$&$C_{2\alpha}$\\
			2&$C_{2\alpha}, m_{(\alpha-\frac{\pi}{2})}$&$C_{2\alpha}$\\
			\multirow{2}*3&$C_{2\alpha}$&$C_{2\alpha}$\\
			&$S^+_{4z},S^-_{4z}$&$S_{4z}$\\
			4-5,51-52,66,74&$C_{2\alpha}, m_\alpha$&$C_{2\alpha}$\\
			\multirow{2}*{6-7}&$C_{2\alpha}, m_\alpha$&$C_{2\alpha}$\\
			&$S^+_{4z},S^-_{4z},C^+_{4z},C^-_{4z}$&$S_{4z}$\\
			8-10,53-54,68&$C_{2\alpha},C^{2\alpha}_z$&$C_{2\alpha}$\\
			11-13&$C_{2\alpha},S^{2(\alpha-\frac{\pi}{2})}_z(\alpha\neq0;\ \tau_o\notin\{G,A,B,C\}(\{G,A,B,C,G',A',B',C'\}for\ 13)$ when  $\alpha=\frac{\pi}{2})$&$C_{2\alpha}$\\
			\multirow{2}*{14-18}&$C_{2\alpha},C^{2\alpha}_z,m_{(\alpha+\frac{\pi}{2})},S^{2(\alpha-\frac{\pi}{2})}_z$&\multirow{2}*{$C_{2\alpha}$}\\
			&$(\alpha\neq0;\  \tau_o\notin\{G,A,B,C\}(\{G,A,B,C,G',A',B',C'\}for\ 18)$ when  $\alpha=\frac{\pi}{2})$&\\
			\multirow{2}*{19-22}&$C_{2\alpha},C^{2\alpha}_z$&$C_{2\alpha}$\\
			&$S^+_{4z},S^-_{4z},m_{\frac{3\pi}{4}},m_{\frac{\pi}{4}}$&$S_{4z}$\\
			\multirow{2}*{23-26}&$C_{2\alpha},S^{2\alpha}_z(\alpha\neq0,\frac{\pi}{2})$&$C_{2\alpha}$\\
			&$S^+_{4z},S^-_{4z},C^{\frac{3\pi}{4}}_2,C^{\frac{\pi}{4}}_2$&$S_{4z}$\\
			\multirow{2}*{27-36}&$C_{2\alpha},C^{2(\alpha-\frac{\pi}{2})}_z,S^{2(\alpha-\frac{\pi}{2})}_z,m_\alpha$&\multirow{2}*{$C_{2\alpha}$}\\
			&$(\alpha\neq0;\ \tau_o\notin\{G,A,B,C\}(\{G,A,B,C,G',A',B',C'\}for\ 35-36)$ when  $\alpha=\frac{\pi}{2})$&\\
			\multirow{2}*{37-48}&$C_{2\alpha},C^{2\alpha}_z,m_\alpha,S^{2\alpha}_z(\alpha\neq0,\frac{\pi}{2})$&$C_{2\alpha}$\\
			&$S^+_{4z},S^-_{4z},C^+_{4z},C^-_{4z},m_{\frac{3\pi}{4}},m_{\frac{\pi}{4}},C^{\frac{3\pi}{4}}_2,C^{\frac{\pi}{4}}_2$&$S_{4z}$\\
			\multirow{2}*{50}&$C_{2\alpha},m_{(\alpha+\frac{\pi}{4})}$&$C_{2\alpha}$\\
			&$E,C_{2z},S^+_{4z},S^-_{4z}$&$S_{4z}$\\
			55-56&$C_{2\alpha},S^{2\alpha}_z(\alpha\neq0,\pm\frac{\pi}{4},\frac{\pi}{2})$&$C_{2\alpha}$\\
			\multirow{2}*{57-58}&$C_{2\alpha},m_{(\alpha+\frac{\pi}{4})},C^{2\alpha}_z,S^{2(\alpha+\frac{\pi}{4})}_z(\alpha\neq\pm\frac{\pi}{4})$&$C_{2\alpha}$\\
			&$E,C_{2z},S^+_{4z},S^-_{4z},C_{2x},C_{2y},m_{110},m_{1\overline{1}0}$&$S_{4z}$\\
			\multirow{2}*{59-60}&$C_{2\alpha},m_{(\alpha+\frac{\pi}{4})},S^{2\alpha}_z,C^{2(\alpha-\frac{\pi}{4})}_z(\alpha\neq0,\frac{\pi}{2})$&$C_{2\alpha}$\\
			&$E,C_{2z},S^+_{4z},S^-_{4z},C^{110}_2,C^{1\overline{1}0}_2,m_{x},m_{y}$&$S_{4z}$\\
			61-64&$C_{2\alpha},S^{2\alpha}_z,C^{2\alpha}_z,m_\alpha(\alpha\neq0,\pm\frac{\pi}{4},\frac{\pi}{2})$&$C_{2\alpha}$\\
			67&$C_{2\alpha},C^{2(\alpha-\frac{\pi}{2})}_z$&$C_{2\alpha}$\\
			69&$C_{2\alpha},S^{2(\alpha-\frac{\pi}{2})}_z(\alpha\neq0,\pm\frac{\pi}{3};\ \tau_o\notin\{G,A,B,C\}$ when $\alpha=\frac{\pi}{2},\pm\frac{\pi}{6})$&$C_{2\alpha}$\\
			70&$C_{2\alpha},S^{2\alpha}_z(\alpha\neq\frac{\pi}{2},\pm\frac{\pi}{6};\ \tau_o\notin\{G,A,B,C\}$ when $\alpha=0,\pm\frac{\pi}{3})$&$C_{2\alpha}$\\
			71&$C_{2\alpha},C^{2(\alpha-\frac{\pi}{2})}_z,m_{(\alpha+\frac{\pi}{6})},S^{2\alpha}_z(\alpha\neq\frac{\pi}{2},\pm\frac{\pi}{6};\ \tau_o\notin\{G,A,B,C\}$ when $\alpha=0,\pm\frac{\pi}{3})$&$C_{2\alpha}$\\
			72&$C_{2\alpha},S^{2(\alpha-\frac{\pi}{2})}_z,m_{(\alpha+\frac{\pi}{6})},C^{2\alpha}_z(\alpha\neq0,\pm\frac{\pi}{3};\ \tau_o\notin\{G,A,B,C\}$ when $\alpha=\frac{\pi}{2},\pm\frac{\pi}{6})$&$C_{2\alpha}$\\
			\multirow{2}*{73}&$C_{2\alpha}$&$C_{2\alpha}$\\
			&$S^+_{4z},S^-_{4z},S^{\frac{\pi}{6}}_z,S^{\frac{5\pi}{6}}_z,S^{-\frac{\pi}{6}}_z,S^{-\frac{5\pi}{6}}_z$&$S_{4z}$\\
			\multirow{2}*{75}&$C_{2\alpha},m_\alpha$&$C_{2\alpha}$\\
			&$S^+_{4z},S^-_{4z},S^{\frac{\pi}{6}}_z,S^{\frac{5\pi}{6}}_z,S^{-\frac{\pi}{6}}_z,S^{-\frac{5\pi}{6}}_z,C^+_{4z},C^-_{4z},C^{\frac{\pi}{6}}_z,C^{\frac{5\pi}{6}}_z,C^{-\frac{\pi}{6}}_z,C^{-\frac{5\pi}{6}}_z$&$S_{4z}$\\
			\multirow{2}*{76}&$C_{2\alpha},C^{2\alpha}_z$&$C_{2\alpha}$\\
			&$S^+_{4z},S^-_{4z},S^{\frac{\pi}{6}}_z,S^{\frac{5\pi}{6}}_z,S^{-\frac{\pi}{6}}_z,S^{-\frac{5\pi}{6}}_z,m_{\frac{\pi}{12}},m_{\frac{\pi}{4}},m_{\frac{5\pi}{12}},m_{\frac{7\pi}{12}},m_{\frac{3\pi}{4}},m_{\frac{11\pi}{12}}$&$S_{4z}$\\
			\multirow{2}*{77}&$C_{2\alpha},S^{2\alpha}_z(\alpha\neq0,\frac{\pi}{2},\pm\frac{\pi}{3},\pm\frac{\pi}{6})$&$C_{2\alpha}$\\
			&$S^+_{4z},S^-_{4z},S^{\frac{\pi}{6}}_z,S^{\frac{5\pi}{6}}_z,S^{-\frac{\pi}{6}}_z,S^{-\frac{5\pi}{6}}_z,C^{\frac{\pi}{12}}_2,C^{\frac{\pi}{4}}_2,C^{\frac{5\pi}{12}}_2,C^{\frac{7\pi}{12}}_2,C^{\frac{3\pi}{4}}_2,C^{\frac{11\pi}{12}}_2$&$S_{4z}$\\
			78&$C_{2\alpha},m_\alpha,S^{2(\alpha-\frac{\pi}{2})}_z,C^{2(\alpha-\frac{\pi}{2})}_z(\alpha\neq0,\pm\frac{\pi}{3};\ \tau_o\notin\{G,A,B,C\}$ when $\alpha=\frac{\pi}{2},\pm\frac{\pi}{6}))$&$C_{2\alpha}$\\
			79&$C_{2\alpha},m_\alpha,S^{2\alpha}_z,C^{2\alpha}_z(\alpha\neq\frac{\pi}{2},\pm\frac{\pi}{6};\ \tau_o\notin\{G,A,B,C\}$ when $\alpha=0,\pm\frac{\pi}{3})$&$C_{2\alpha}$\\
			\multirow{2}*{80}&$C_{2\alpha},m_\alpha,S^{2\alpha}_z,C^{2\alpha}_z(\alpha\neq0,\frac{\pi}{2},\pm\frac{\pi}{6},\pm\frac{\pi}{3})$&$C_{2\alpha}$\\
			&$C_{2\alpha},m_\alpha,S^{2\alpha}_z,C^{2\alpha}_z(\alpha=\frac{n\pi}{12},n=1,3,5,7,9,11)$&$S_{4z}$\\
		\end{tabular}
	\end{ruledtabular}
\end{table*}

We now focus on the translation part of the stacking operation, i.e., the solutions of Equation (\ref{eq:10}). Because $E+\overline{E}$ is zero, there is no constraint on $\tau_{o}$ for $\overline{E}$, which implies that any sliding operation cannot break the inversion symmetry in the bilayer. For $m_z$, the solutions are $\tau_{o} = \{G, A,B,C\}$, where $G=(0,0)$, $A=(\frac{1}{2},0)$, $B=(0,\frac{1}{2})$, $C=(\frac{1}{2},\frac{1}{2})$, employing lattice vectors as basis vectors. $\tau_{o}=G$ means no translation. If the system is a centered rectangular lattice, $\tau_{o} = \{G, A,B,C,G',A',B',C'\}$ for $m_z$, where $G'=(\frac{1}{4},\frac{1}{4})$, $A'=(\frac{3}{4},\frac{1}{4})$, $B'=(\frac{1}{4},\frac{3}{4})$, $G'=(\frac{3}{4},\frac{3}{4})$. For $C_{2\alpha}$, translation of arbitrary length perpendicular to the vector (cos($\alpha$), sin($\alpha$), 0), as well as translations by half of the lattice vector in the direction of vector (cos($\alpha$), sin($\alpha$), 0), and their combinations, are all permitted. For $S_{3z}$, $\tau_{o}$ is only equal to $G$, which implies that any sliding operation can break it. For $S_{6z}$, $\tau_{o}=\{G,M,N\}$ are obtained, where $M=(\frac{1}{3},\frac{2}{3})$, $N=(\frac{2}{3},\frac{1}{3})$. For $S_{4z}$, the solutions of $\tau_{o}$ does not affect the results of this article, and different lattices have different solutions. Therefore, we provide only  a general solution:
\begin{equation}
	\left\{\label{eq:12}
	\begin{aligned}
		\tau_{ox}&=\frac{1}{2}(\tau_{s0x}+\tau_{s0y})\\
		\tau_{oy}&=\frac{1}{2}(\tau_{s0y}-\tau_{s0x})
	\end{aligned}
	\right.
\end{equation}
$\tau_{ox}$ and $\tau_{oy}$ are the \emph{x}-component and \emph{y}-component of $\tau_{o}$, respectively. One of the lattice vectors of the conventional cell is chosen to serve as the \emph{x}-axis. 

By excluding the stacking operations that generate $\{\overline{E},m_z\}$ in the bilayer from those that generate $\{C_{2\alpha},S_{nz}\}$ in the bilayer for each layer group, all BSAAs are obtained, as listed in Table \ref{table1}. The second column of Table \ref{table1} lists all  stacking operations to form BSAA, for a ferromagnetic monolayer with layer group listed in the first column. The third column lists the symmetry exchanging the positions of two ferromagnetic layers in the bilayer under the stacking operation listed in the second column.

Based on Table \ref{table1}, we derive the following three conclusions. Firstly, only 17 layer groups can realize intrinsic A-type altermagnetism, including layer group numbers 8-10, 19-22, 50, 53-54,57-60,67-68,76. We regard the bilayer as intrinsic A-type altermagnet, if the top layer is obtained by translating the bottom layer along the z-direction without any other operations and it possesses A-type altermagnetism. Because the stacking operation is $\hat{\tau}_z\{0|E\}$, the symmetry of monolayer is the same as bilayer, i.e., they possess the same layer group. Therefore, this conclusion has two implications. The first one is that only ferromagnetic monolayers with these layer groups can form BSAA through $\hat{\tau}_z\{0|E\}$, without any other operations. The second is that all 2D A-type altermagnets must belong to these 17 layer groups. Obviously, not all 2D A-type altermagnets can be regarded as being formed by two sublattices through stacking operation $\hat{\tau}_z\{0|E\}$. This is because the minimum energy structure of 2D materials may require sliding, twisting or other adjustments to achieve. However, it will not affect our result. If the minimum energy structure possesses 2D A-type altermagnetism, its layer group must belong to these 17 layer groups. In other words, A-type antiferromagnetic bilayers that belong to these 17 layer groups will exhibit altermagnetism. Furthermore, the layer group of all BSAAs belong to these 17 layer groups. 

Secondly, it is impossible to connect the two sublattices of BSAA using $S_{3z}$ or $S_{6z}$. This is because the emergency of $S_{6z}$ is always accompanied by inversion symmetry, and  the emergency of $S_{3z}$ is always accompanied by $m_z$ symmetry. Moreover, the inversion symmetry in the bilayer cannot be broken by sliding operation, and any sliding operation will break $S_{3z}$ symmetry. In fact, this result is consistent with our previous work\cite{zeng2024descriptiontwodimensionalaltermagnetism}. All 2D altermagnets are described using seven spin layer groups. There are not spin-group symmetry $[C_2||S_{3z}]$ and $[C_2||S_{6z}]$ in all spin layer groups. It implies that it is impossible to connect the two sublattices of all 2D altermagnets using $S_{3z}$ or $S_{6z}$. Therefore, this conclusion is also applicable to all 2D altermagnets. 

Finally, $C_{2\alpha}$, where $\alpha$ is an arbitrary angle within the range from 0 to $\pi$, is a general stacking operation to generate BSAA for an arbitrary monolayer.  Meanwhile, we notice that not all A-type antiferromagnetic bilayers can generate altermagnetism only through twisting. From a stacking perspective, the twisting can be regarded as stacking operation $C_z^{2\alpha}$. Based on Table \ref{table1}, it is not a general stacking operation to generate BSAA.

\begin{figure*}
	
	\centering
	\includegraphics[width=\linewidth]{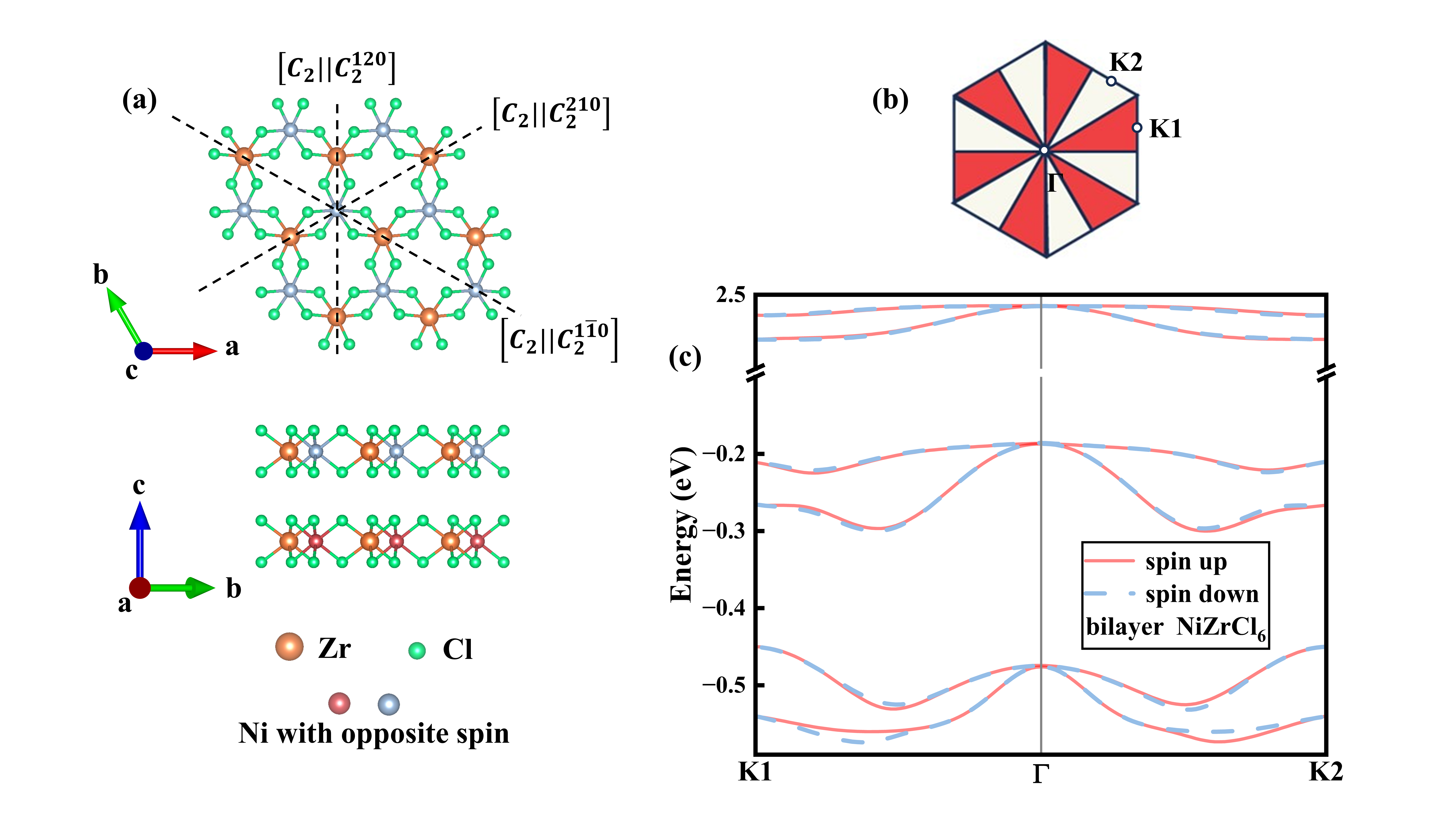}
	\caption{\label{fig:figure1}The crystal structure and nonrelativistic band structure of bilayer NiZrCl$_6$. (a) The crystal structure of bilayer NiZrCl$_6$, where red and blue colors represent the opposite spin sublattices. Three spin-group symmetries, which interconnect two ferromagnetic layers, have been shown in (a). (b) The Brillouin zone of bilayer NiZrCl$_6$, in which the different colors represent opposite spins, and high-symmetry points used to calculate band structure. (c) The band structure of bilayer NiZrCl$_6$ without SOC along \textbf{k}-path K1-$\Gamma$-K2 is spin-splitting, where red solid lines and blue dotted lines represent the opposite spin channels. The spin-splitting band structures in (c) illustrate that bilayer NiZrCl$_6$ exhibits altermagnetism.}
	
\end{figure*}

\section{\label{sec:level3} BILAYER WITH INTRINSIC A-TYPE ALTERMAGNETISM}
According to the first conclusion of the last section, a bilayer characterized by intralayer ferromagnetic coupling and interlayer antiferromagnetic coupling, and belonging to one of 17 layer groups, exhibits A-type altermagnetism. Based on these layer groups, we search for BSAA in van der Waals Bilayer Database (BiDB)\cite{pakdel2024high} as the example of BSAA. Firstly, we select out materials with these 17 layer groups. Then, we examine their magnetic coupling properties. If the magnetic configuration is A-type, we stop our search. A comprehensive search of the database is not conducted.

Based on the above approach, we predict that bilayer NiZrCl$_6$, in which the top layer is obtained by applying $\tau_z$ on the bottom layer, exhibits intrinsic A-type altermagnetism. The crystal structure of bilayer NiZrCl$_6$ is schematically illustrated in FIG. \ref{fig:figure1}(a), with red and blue colors representing the sublattices with opposite spins in A-type altermagnetism. This structure is stable, according to the data in BiDB. The monolayer NiZrCl$_6$ exhibits ferromagnetism, which has been reported previously\cite{10.1063/5.0158822}. Meanwhile, the layer group number of monolayer NiZrCl$_6$ is 67, which surely belongs to the 17 layer groups. Therefore, as long as the interlayer magnetic coupling is antiferromagnetic, the bilayer NiZrCl$_6$ is BSAA, according to our theory. Using the first-principles calculation, we compare the energies of different magnetic configurations, ferromagnetic and A-type altermagnetic. The energy of A-type altermagnetic state is 0.25 meV/f.u. lower than that of the ferromagnetic state, which indicates that its magnetic ground state is A-type altermagnetism.

Through symmetry analysis, the spin layer group of bilayer NiZrCl$_6$ is determined as	$^{1}\overline{3}$$^{2}m$. The sublattices are related by three 2-fold rotation symmetries along the rotational axis within \emph{xy}-plane, as illustrated in FIG. \ref{fig:figure1}(a). Consequently, the spin-momentum locking of bilayer NiZrCl$_6$ is determined by the symmetries $[C_2||\{C_2^{1\overline{1}0},C_2^{120},C_2^{210}\}]$, as illustrated in FIG. \ref{fig:figure1}(b). The \textbf{k}-paths, related by $C_2^{1\overline{1}0}$  ($C_2^{120}$, $C_2^{210}$) but not separated by a reciprocal lattice vector, have opposite spin signs. FIG. \ref{fig:figure1}(c) shows the DFT calculated band structure without SOC and the employed \textbf{k}-path is shown in FIG. \ref{fig:figure1}(b). The spin-splitting band structure along K1-$\Gamma$-K2 and the zero net magnetic moment imply that bilayer NiZrCl$_6$ is an altermagnet. This result is consistent with our theory. The bilayer NiZrCl$_6$ corresponds to the case where layer group is \#67 and $\alpha = \frac{\pi}{2}$ in Table \ref{table1}. 

\section{\label{sec:level4} TWISTED-BILAYER WITH A-TYPE ALTERMAGNETISM}
Twisted-bilayer systems with A-type altermagnetism have been researched\cite{PhysRevLett.130.046401,liu2024twistedmagneticvander}. It is noteworthy that a twisted-bilayer can be regarded as two single layers in which the top layer is rotated and then stacked onto the bottom layer. In this article, it corresponds to the stacking operation $C_z^{2\alpha}$. In this case, $\alpha$ is the twisting angle. Therefore, we posit that twisted-bilayer with A-type altermagnetism is also BSAA. The twisted-bilayer NiCl$_2$ has been predicted as A-type altermagnet in previous report\cite{PhysRevLett.130.046401}. We here take it as an example of BSAA to verify our result.

The layer group number of the monolayer 1T-phase NiCl$_2$ is 72. Based on Table \ref{table1}, to achieve A-type altermagnetism, $\alpha$=\{$0, \pm\frac{\pi}{3}$\} are forbidden, but the Moir\'{e} angle of the structure, including $13.17^\circ, 21.79^\circ, 27.79^\circ, 32.21^\circ, 38.21^\circ$ and $46.83^\circ$, are permitted. This is because the inversion symmetry is broken by twisting with a Moir\'{e} angle, but preserved by twisting with $\alpha$=\{$0, \pm\frac{\pi}{3}$\}. This result consistent with the previous research\cite{PhysRevLett.130.046401}. It is noteworthy that when twisting angle is equal to $\frac{\pi}{2}$ or $\pm\frac{\pi}{6}$, the mirror symmetry through \emph{xy}-plane, $m_z$, is generated in the twisted-bilayer NiCl$_2$. Therefore, to achieve A-type altermagnetism, the extra sliding is necessary for twisted-bilayer NiCl$_2$ with a twisting angle of $\frac{\pi}{2}$ or $\pm\frac{\pi}{6}$, as listed in Table \ref{table1}.

\begin{figure*}
	
	\centering
	\includegraphics[width=\linewidth]{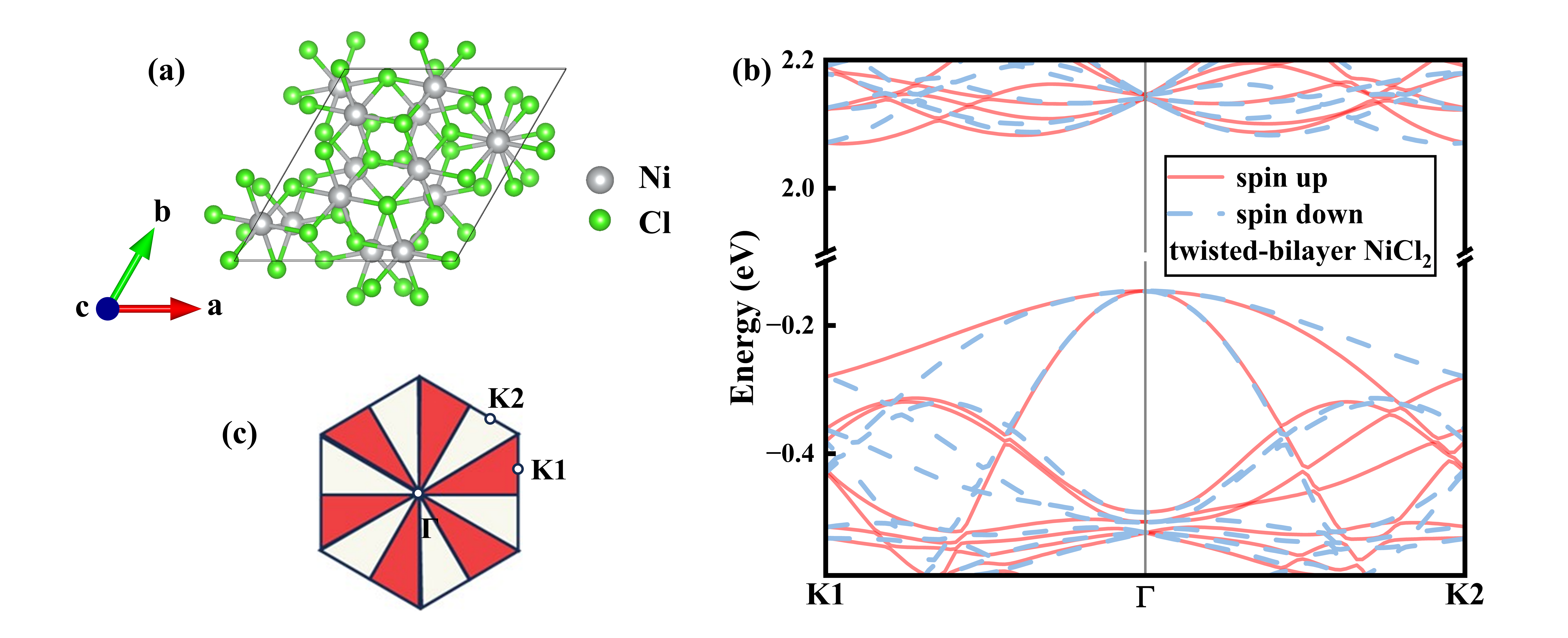}
	\caption{\label{fig:figure2}The crystal structure and nonrelativistic band structure of twisted-bilayer NiCl$_2$ with a twisting angle of $21.79^\circ$. (a) The crystal structure of twisted-bilayer NiCl$_2$. (b) The band structure of twisted-bilayer NiCl$_2$ without SOC along \textbf{k}-path K1-$\Gamma$-K2 is spin-splitting, where red solid lines and blue dotted lines represent the opposite spin channels. (c) The Brillouin zone of twisted-bilayer NiCl$_2$, in which the different colors represent opposite spins, and high-symmetry points used to calculate band structure. The spin-splitting band structures in (b) illustrate that twisted-bilayer NiCl$_2$ exhibits altermagnetism.}
	
\end{figure*}

We now focus on the twisted-bilayer NiCl$_2$ with a twisting angle of $21.79^\circ$. The crystal structure is schematically illustrated in FIG. \ref{fig:figure2}(a). Through symmetry analysis, we determine that its spin layer group is $^{1}\overline{3}$$^{2}m$ and its layer group number is 67. As mentioned earlier, for the material with 2D A-type altermagnetism, it must be one of the 17 layer groups. Interestingly, the twisted-bilayer NiCl$_2$ has the same spin layer group with NiZrCl$_6$, which has been discussed in Section \ref{sec:level3}. Consequently, they have the same spin-momentum locking. We use the same \textbf{k}-path, as illusctrated in FIG. \ref{fig:figure2}(c), to calculate band structure. FIG. \ref{fig:figure2}(b) shows the DFT calculated band structure without SOC, which demonstrates that it clearly exhibits A-type altermagnetism.

\section{\label{sec:level5} CONCLUSION}
In summary, we propose a new concept of bilayer stacking A-type altermagnet(BSAA), deduce all BSAAs for all layer groups, and provide two distinct examples of BSAAs. Our research has significantly broadened the range of candidate materials for 2D altermagnets. Based on all BSAAs we have derived, we obtain three intriguing findings: 

(\romannumeral1) Only 17 layer groups can realize intrinsic A-type altermagnetism. All 2D A-type altermagnets must belong to these 17 layer groups, which has never been reported previously. Of course, all BSAAs listed in Table \ref{table1} belong to these 17 layer groups. A-type antiferromagnetic bilayers that belong to these 17 layer groups will exhibit altermagnetism. 

(\romannumeral2) It is impossible to connect the two sublattices of BSAA using $S_{3z}$ or $S_{6z}$. Combining the past research\cite{zeng2024descriptiontwodimensionalaltermagnetism}, we find that this conclusion is applicable to all 2D altermagnets. 

(\romannumeral3) $C_{2\alpha}$ is a general stacking operation to generate BSAA for an arbitrary monolayer. 

Based on symmetry analysis and DFT calculation, we predict a new 2D altermagnet, the bilayer NiZrCl$_6$, which has never been reported in the past. Meanwhile, we notice that the previously reported twisted-bilayer A-type altermagnet is contained within BSAA. We use twisted-bilayer NiCl$_2$ as the second example of BSAA.

\begin{acknowledgments}
	This work is financially supported by National Natural Science Foundation of China (Grant No. 12074126). The computer times at the High Performance Computational center at South China University of Technology are gratefully acknowledged.
\end{acknowledgments}

\appendix
\section{COMPUTATIONAL DETAILS}
All DFT calculations were performed using the Vienna ab initio Simulation Package (VASP)\cite{kresse1996efficiency,kresse1996efficient},  employing the projector-augmented wave (PAW) method\cite{kresse1999ultrasoft} based on density functional theory. For the exchange-correlation functional, we use the generalized gradient approximation (GGA) with Perdew-Burke-Ernzerhof (PBE) functional\cite{perdew1996generalized}, along with Hubbard U correction\cite{PhysRevB.57.1505}. The parameters for calculation are listed in Table \ref{tab:table2}. A cut-off energy of 600 eV was set for the plane wave basis. The structure was relaxed until the forces on atoms were below 0.01 eV/\AA\ and the convergence criteria was $1\times10^{-7}$ eV for the energy difference in electronic self-consistent calculation. A vacuum of 18\AA\ was constructed perpendicular to the material plane. We utilize the DFT-D3 approach\cite{https://doi.org/10.1002/jcc.21759} to correct van der Waals (vdW) interaction. The SOC effect was not considered in calculations.
\begin{table}[H]
	\caption{\label{tab:table2}The parameters for calculation. The lattice vector of NiCl$_2$ listed in the third column, \emph{a}/\emph{b}, is lattice vector of the supperlattice of bilayer NiCl$_2$. The interlayer distance represents the distance between two layers in bilayer. }
	\begin{ruledtabular}
		\begin{tabular}{ccccc}
			\textrm{material}&
			\textrm{Hubbard U (eV)}&
			\textrm{\emph{a}/\emph{b} (\AA)}&
			\textrm{k-mesh}&
			\textrm{interlayer distance (\AA)}\\
			\hline 
			NiZrCl$_6$ & 3 & 6.20 & 14$\times$14$\times$1 & 3.18\\
			NiCl$_2$ & 3 & 9.14 & 6$\times$6$\times$1& 3.33 \\
		\end{tabular}
	\end{ruledtabular}
\end{table}

\bibliography{ref}

\end{document}